# Revisiting Abstractions for Software Architecture and Tools to Support Them

Mary Shaw, *Life Fellow, IEEE*, Daniel V. Klein, and Theodore L. Ross

*Abstract*—The mid-1990s saw the design of programming languages for software architectures, which define the high-level aspects of software systems including how code components were composed to form full systems. Our paper *Abstractions for Software Architecture and Tools to Support Them* presented a conceptual view of software architecture based on abstractions used in practice to organize software systems, a language that supported these abstractions, and a prototype implementation of this language [1]. By invitation, we reflect on the paper's principal ideas about system-level abstractions, place the work in a historical context of steadily increasing abstraction power in software development languages and infrastructure, and reflect on how progress since the paper's 1995 publication has been influenced, directly or indirectly, by this work. We describe current manifestations of architectural ideas and current challenges. We suggest how the strategy we used to identify and reify architectural abstractions may apply to current opportunities.

## I. Abstractions and Tools for Software Architecture

SOFTWARE architecture addresses high-level aspects of software systems such as the overall organization, the decomposition into different types of components, the assignment of functionality to components, and the various ways the components interact. In the mid-1990s, software engineering experienced a transition from module interconnection languages that treated components uniformly to software architecture languages that supported a richer and more nuanced set of overall architectural styles, together with abstractions for different types of components and the ways they interact. The language development was accompanied by design guidance about how to select a software architecture appropriate to a particular problem.

In *Abstractions for Software Architecture and Tools to Support Them* we made the case for first-class language support for architectural descriptions [1]. This included the overall "style" of system organization (such as pipes and filters or layers). It also distinguished different types of components (such as filters, processes, and data stores) and different types of connectors, or protocols, that govern the interactions of specific types of components (such as pipes, remote procedure calls, or event signaling). We described a textual and visual language, UniCon, and a prototype implementation that could compile descriptions in the language.

The principal ideas were:

- Support for component-level abstractions rather than conventional programming language features
- Identification of commonly-used architectural styles or patterns
- Connectors as important abstractions that define interactions among components
- Distinct types of components and distinct types of connectors, with type-matching rules for how to compose them
- Visual and textual language for describing architectures, supplementing conventional programming languages
- Libraries of developer-defined component and connector types

The essential insight supporting these ideas was that developers have a shared folklore of common useful software organizations and that reifying those patterns—recognizing them, defining them, and supporting them explicitly—would provide benefits not only in software creation but also in reuse, understanding, analysis, and maintenance.

## II. Historical Context

The impact of *Abstractions for Software Architecture and Tools to Support Them* can be understood through the lens of the historical development of abstract concepts in programming languages and software development tools. These constructs were initially language features (e.g., macros, objects), but as time passed they came to encompass system abstractions (e.g., architectural styles, product lines, systems).

For this contextual lens we provide an informal chronology of the emergence of software abstractions into popular use, locating software architecture abstractions on the timeline of language and tool abstractions. Using this abstraction lens and the Redwine-Riddle model of technology maturation [2], we show how the ideas in our paper influenced the maturation of the field of software architecture.

This work was supported in part by the Alan J. Perlis Chair of Computer Science at Carnegie Mellon University. (*Corresponding author: Mary Shaw*).

Mary Shaw is with the School of Computer Science, Carnegie Mellon University, Pittsburgh PA 15217 USA (email: mary.shaw@cs.cmu.edu).

Daniel V. Klein is now at Google DeepMind, Pittsburgh PA 15206 USA. This work was done independently of Google DeepMind. (e-mail: daniel.v.klein@gmail.com).

Theodore L. Ross is with Red Hat Inc Lowell MA 01851 USA (e-mail: ted.ross@verizon.net).

All authors were with Carnegie Mellon when [1] was published in 1995.

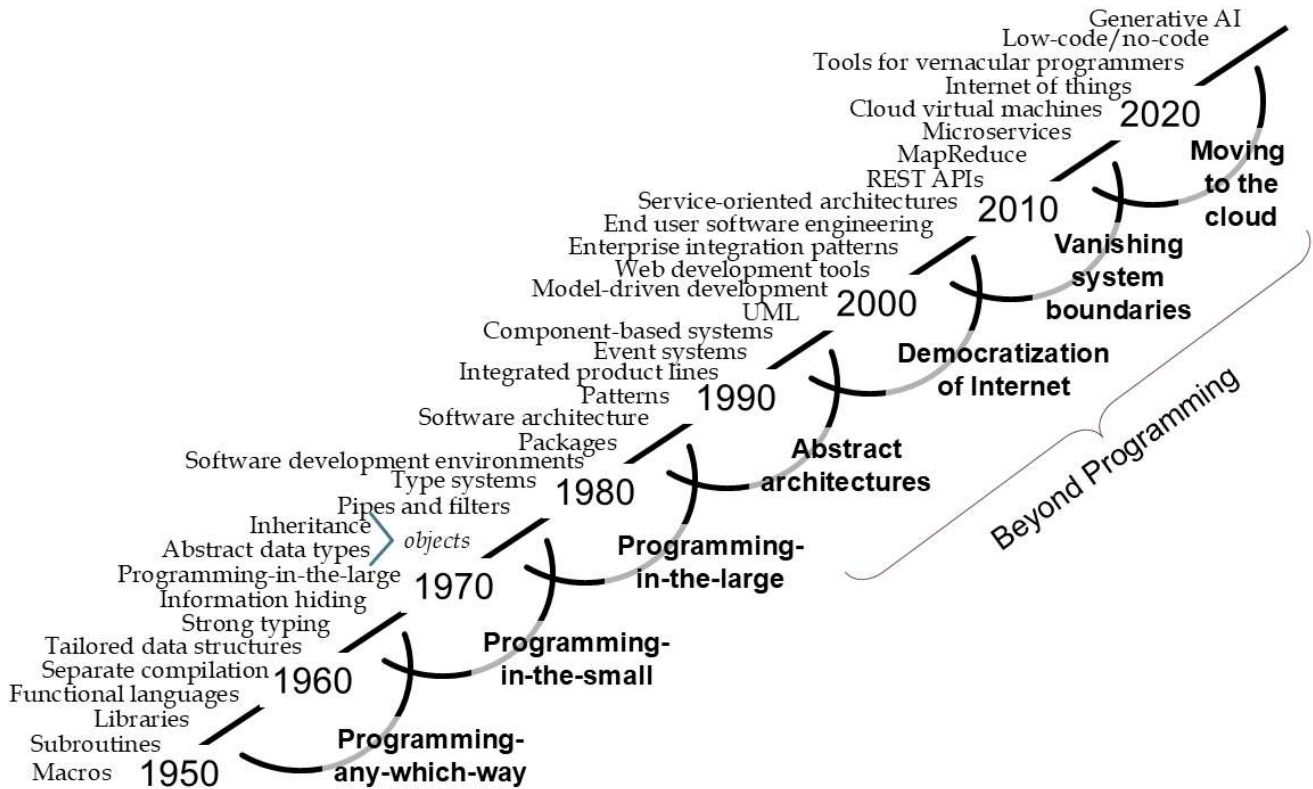

Figure 1. Growth of software development abstractions over time

*A. Abstraction Power in Programming Languages*

We can track progress in software development by the conceptual size of the abstractions we use without "looking inside". Initially these abstractions were programming language features: there was a natural progression in which the abstractions grew larger and more powerful through time [3]. That is, the amount of computation you could invoke with each line of code grew.[1]

Over the years the abstractions of interest outgrew conventional programming languages. For a time they captured ideas about architectures—about how different types of components are composed with different protocols to form systems. As time passed, numerous classes of systems developed their own ecosystems including domain-specific architectures and tooling such as frameworks, development environments, platforms, databases, and development environment [4]. Figure 1 illustrates this abstraction growth over the past ¾ century.

*B. Maturation of Software Architecture*

The abstractions of software architecture were identified during the long-term evolution of software abstractions. The contributions establishing this subfield of software engineering were concentrated in the period from 1989 to the early 20-teens. We turn now to the maturation of software architecture and the role our paper played in that maturation.

We rely on the Redwine-Riddle model of software technology maturation [2]. They identified six phases of maturation along with events or capabilities that indicate transitions between the phases (often with considerable overlap between phases). Their contemporary examples showed 15-20 years (with high variance) from basic research to the point where a technology is ready for popularization. Their six stages of maturity and the events or capabilities that supported a transition to the next stage are:

- *Basic research,* foundations, and new ideas, with a key idea leading to …
- *Concept formation* and refinement of ideas, with a seminal paper or system leading to …
- *Development and extension* through experimentation and refinement, with usable capability leading to …
- *Internal exploration* by developers on their own real problems, with emerging outsider use leading to …
- *External exploration* by a broader user group, with production quality commercial support leading to …
- *Popularization* and propagation through a user community

This provides a framework for reflecting on the maturation of software architecture [5].

*Basic research (1970s to 1990s)*

Software architecture stands on the shoulders of module interconnection languages, which emerged in the 1970s [6][7]. Object-oriented programming rose to prominence in the late 1970s and 1980s, though it was not appropriate for all software. In the 1990s UML provided a suite of notations for describing certain system-level properties of object-oriented models [8]. Programmers were using informal shared idioms to describe many system organizations that were not object-oriented. Explicit architectures were being developed in specific domains.

---

[1] This discussion is intended to be evocative, showing the increasing power available to the software developer. It is informal and incomplete. A scholarly history would include much more detail, supported by proper citations; it would have more and different items in a somewhat different order.

*Concept formation (1989-1996)*

Dissatisfaction with the object model as the sole model for system structure set the stage for research in software architecture beginning in the late 1980s. We began cataloging the informal architectural idioms into a more systematic form [9], followed by identifying concepts of architecture [10][11][12] and architectural style [13][14]. A key insight was recognizing that the interactions among components were of distinct types, but these abstractions were obscured by their implementations as specific sequences of procedure calls on system routines [15]. Our "Abstractions and Tools…" paper [1] brought many of these ideas together in an architecture description language, supported by a prototype implementation. This was one of seven papers in a special issue of *IEEE TSE* on software architecture [16].

*Development and extension (1995-2000)*

The next few years refined the concepts. Patterns were systematized and classified [17], along with work on interoperability and integration [18]. Several architectural description languages (ADLs) were defined [19]. The institutions of a discipline started emerging, including conferences and textbooks [20].

*Internal Exploration (1996-2003)*

Architectural ideas clarified existing practices, so explicit attention to architecture in design followed quickly after the original expositions. This period saw the use of styles and patterns as design guides, attention to quality attributes, and analysis and evaluation techniques [21][22][23].

*External Exploration (1999-2010)*

Adoption of architectural approaches to design spread quickly after accessible books explained the ideas, early experience showed promise, and development process frameworks provided mature support through management processes, tools, use cases, and sample models [24]. Companies started developing their own end-to-end architectures to normalize their product lines.

Software architecture also thrived as a research area. In 2005 a CiteSeer search found about 750 distinct books and papers with "software architecture" in the title; these had over 500 distinct lead authors; 75% were published after 1994 [25].

By 2010 an NRC report [26] identified software architecture as one of three critical areas for software. The title "software architect" was sought after: CNN Money identified this as the best job in America [27].

*Popularization (2008-)*

As production-quality frameworks, processes, and other tools emerged, architectural support became commercialized, with frameworks often serving as the carriers of the architecture. Architects became recognized as senior technical leaders. In 2014, the architecture definition process was included in the drafts of ISO/IEC/IEEE standards for software and system life cycle processes [28][29].

Vernacular programmers—people who are not professionally trained developers but who develop software for their own needs— also gained better support. For example, architectural patterns for web-based systems were supported by ecosystems of frameworks, development environments, platforms, databases, and development environments [4].

*Beyond simple architectures*

This cycle of the development of architectures for localized systems established the concept of architecture as part of software design. This set the stage for addressing more complex architectures for highly heterogeneous distributed systems and for specific domains, each of which has its own maturation cycle. We turn to some of these opportunities, which matured in their own time, in Section III.

### III CURRENT ARCHITECTURAL ABSTRACTIONS AND TOOLS

When we wrote "Abstractions and Tools…" [1], we expected that architecture description languages would be used in the same way as programming languages: The software developer would deliberately select an architecture for the problem at hand and write a complete architectural description in the ADL; this architecture-level code would serve as the authoritative definition. It would be compiled by a super-make-like system, and it would not be mere documentation because it would be the master compilation source.

As support for software architecture evolved, it wound up at a very different point in the generality-power tradeoff space. Abstractions for specific domains have found their way into the programming languages for those domains. Frameworks, platforms, scripting languages and other tools support specific architectures in richer ways. These integrate into workflows and preempt the architectural decisions. Indeed, a general-purpose programming language plays only a minor role in the ecosystem for web-based systems mentioned above. Similarly, designing support for richer modern architectures requires integrating support with support for other capabilities. This section discusses some current work shaped by these forces.

*A Domain-Specific Architectural Support in Software Tools and Languages*

Our original paper helped to define the field of software architecture and shape the first phase of its evolution. Since the initial success, the field has evolved in a way consistent with the Redwine and Riddle model. Today, few developers need to concern themselves with the conceptual complexity of the architecture they are creating, because much of the effort in creating a system has been abstracted away into language and infrastructure features. Now even vernacular programmers can create arbitrarily complex architectures with ease. This may be accomplished with first-class language features (such as GoRoutines for parallelism and scheduling [30]) or libraries and operator overloading (such as Dataflow [31] and Apache Beam [32]). Dataflow provides a fully managed service for executing Apache Beam pipelines, offering features like autoscaling, dynamic work rebalancing, and a managed execution environment, and Beam is an open source unified programming model to define and execute data processing pipelines, including ETL, batch and stream (continuous) processing.

While Unix pipes and filters embodied one particular architecture style, Dataflow/Beam extends this by hiding multiple different styles and extended deployment and behavioral semantics behind a similar syntax. Using Dataflow/Beam as an example, almost all a programmer must do to create a pipeline is to define routines A, B, and C (using a flexible but standardized interface convention), and connect them with the overloaded '|' operator.

```
input | A() | B() | C() | output
```

When the program executes, Dataflow/Beam parallelizes the execution of A, B, and C based on the dimensionality of the input, so that the running program might at any time have tens of thousands of workers spread across a datacenter at any given time (or it can parallelize the job across the limited number of cores on a desktop). The programmer barely needs to be concerned with any details beyond the flow of data between stages. No longer does the programmer need to consider whether the flow of data should be implemented with Unix pipes, RPCs, a pub/sub system or something else. While preserving the traditional Unix syntax, modern Dataflow/Beam provides parallelism, rebalancing, communication, failsafe, and retry.

Similar examples exist for databases. With Google BigQuery [33], a simple SQL query can result in thousands of workers scanning petabytes of data across a datacenter, without the query writer doing anything outside of their ordinary SQL workflow.

Historically, this has always been the trend. When our original paper was written, we needed concrete examples to demonstrate Unicon, and so the implementation included some rather clever code for implementing pipelines that forked, joined, or were even circular. The special knowledge to do this was available 15 years earlier, in the source-code for Unix utilities. By 1992, few people outside of the kernel development community had ever used this power (it involves figuring out how the `dup()`, `pipe()`, `fork()`, `exec()` and `wait()` system calls are used by the shell to create a simple pipeline). Today Unix file redirection and pipelines "just work" in the shell, and fewer people learn the details, relying instead on larger, purpose built tools.

In the machine shop, using a lathe is a dying art. Nowadays most people learn CNC or 3D printing and modeling, and the machines/programs do the work for them. Similarly, we observe that many kinds of architectural design have been subsumed by domain-specific ecosystems. In other words, we have successfully abstracted the architectures into generalized language features so that for most users, they "just work", and the details are largely irrelevant.

*B. Ongoing Challenges in Heterogeneous Highly Distributed Systems*

When our original paper was written, the component interconnections of interest were primarily between components in close proximity. This includes separate objects linked together into an executable image and separate executables on a host communicating via named pipes or remote procedure calls.

In the time since the publication of our paper, software architectures have become far more widely distributed. These distributed software systems use data networks, almost exclusively TCP/IP, as the communication substrate.

The fact that TCP/IP now underlies much of the connecting glue in distributed software architectures introduces challenges not previously in play that work against establishing clean abstractions for distributed architectures. The primary challenge introduced by networks is the way they handle addressing. Unlike with a named pipe, where an explicit source-sink relationship can be established between components, network addressing provides a way to discover, locate, and reach only the destination or server role in the relationship. There is no formality with regard to the identity of the component in the source or client role. Further, this discussion only addresses the connectivity aspect of the network, and not the semantic characteristics of the data flowing over that connection—but connectivity is a prerequisite for semantics.

There are a number of degrees of separation between distributed software components that further complicate the problem: Addressing is straightforward for components in a data center sharing a common network, or containerized workloads being orchestrated by a platform like Kubernetes. However, as we progress into the "moving to the cloud" era, we see inter-component communication that needs to cross firewall boundaries; between public and private subnetworks separated using Network Address Translation (NAT); between disparate private subnetworks that have overlapping address spaces (common in edge computing) or that exclusively use different IP versions (IPv4 vs IPv6). Pure IPv6 environments are prevalent in Asia Pacific countries for example.

The opportunity going forward is to build better abstractions for connectivity that allow software architects to dispense with concern over the complexity of the networks underlying their systems. Work is being done in this area, for example in the open source Skupper project [34].

Once good abstractions for connectivity are in place, the table will be set for the emergence of models and notations for distributed software architecture that can once again handle complex connectors and high-level concepts with as much internal complexity as the components themselves. Such models will then support reasoning about increasingly large distributed systems from a number of different perspectives and areas of concern (security, performance, regulatory compliance, resilience, resource consumption and cost, correctness, etc.).

## V. What's Next/

The enduring idea from *Abstractions for Software Architecture and Tools to Support Them* [1] is that each level of software design has some common patterns that serve developers well. Those abstractions are not, however, always initially obvious; it's necessary to study the ad hoc practice and identify the structure of useful patterns. Reifying those patterns as named abstractions helps software developers reuse the ideas, understand other developers' software better, support systematic reasoning, and avoid re-inventing gratuitously similar structures [35]

We worked through that process for the architecture level of software design: we moved beyond module implementation language to identify specific architectural patterns in common use and created named abstractions that gave them identities, well-defined structure, and a basis for reasoning. That is, we identified and reified common abstractions at the architecture level of design.

We call out the approach of seeking conceptual patterns—not just syntactic patterns—in complex settings in order to find appropriate useful abstractions. We believe that this is a useful strategy, applicable to many areas.

Here are some of our hopes and aspirations for applying this approach going forward:

- *Abstractions for highly complex heterogeneous distributed systems.* These will be predicated upon elevated abstractions for network interconnect and will address additional concerns including security, performance, compliance, cost, and interoperability of different system architectures.
- *Definitions of useful design spaces*. Designers must make decisions that balance multiple, often interacting, desiderata. Expressing the important structure of the problem domain as a design space often makes the consequences of design choices more transparent [36].
- *Abstractions that better match the mental models of vernacular programmers.* Domain-specific languages often serve these developers better than powerful general-purpose languages, and they also need better support for maintenance concerns, software sharing, cloud computing, and deciding whether their software is sufficiently correct [4].
- *Explicit identification of design obligations*. Software languages, architectures, and paradigms often have expectations beyond writing well-formed definitions, for example about null pointers, invariants, or consistent look-and-feel. These obligations are often tacit but essential to satisfactory results. Undocumented obligations may go unnoticed by developers, and they are hard to observe and check. They deserve to become explicit abstractions [37].
- *Using generative AI as a tool to discover useful patterns for abstraction.* Modern AI tools can help find common patterns, though the conceptual abstractions that reify these patterns are often tacit. Generative AI may help find patterns, under careful human supervision [38].

## Acknowledgment

We thank our original co-authors Rob DeLine, Dave Young, and Greg Zelesnik for their contributions to the 1995 paper [1]; we regret that various circumstances did not allow them to engage with this retrospective. We have valued ongoing collaborations with colleagues at Carnegie Mellon who were integral to our progress in software architecture, especially David Garlan and Bradley Schmerl, who also offered helpful comments on this paper. We appreciate George Fairbanks' helpful comments about programming abstractions and thoughtful comments from Marian Petre.

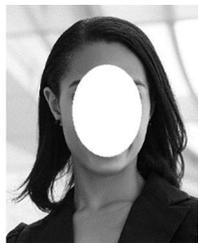

**Mary Shaw** (Life Fellow, IEEE) is the Alan J. Perlis University Professor of Computer Science in the School of Computer Science at Carnegie Mellon University, Pittsburgh PA, where she served as Associate Dean for Professional Programs (1992-1999) and Chief Scientist of the Software Engineering Institute (1984-1997). She earned a PhD in computer science in 1971 from Carnegie Mellon University and a BA *cum laude* in mathematics in 1965 from Rice University, Houston TX.

She has made contributions to an engineering discipline for software through developing data abstraction with verification (with W. Wulf and R. London), establishing software architecture as a major branch of software engineering (with D. Garlan), working on high-level design and design spaces in adaptive systems and systems for vernacular programmers, designing influential and innovative curricula in software engineering and computer science supported by two influential textbooks, and helping to found the Software Engineering Institute at Carnegie Mellon.

Prof. Shaw has received the United States' National Medal of Technology and Innovation, the IEEE Computer Society TCSE's Lifetime Achievement, Distinguished Educator, and Distinguished Women in Software Engineering Awards, the George R. Stibitz Computer & Communications Pioneer Award, the ACM SIGSOFT Outstanding Research Award (with David Garlan), and CSEE&T's Nancy Mead Award for Excellence in Software Engineering Education. At Carnegie Mellon she has received the Robert E. Doherty Award for Sustained Contributions to Excellence in Education and the Allen Newell Award for Research Excellence (with David Garlan and Bradley Schmerl). She is an elected Fellow and Life Member of the ACM, the IEEE, and the American Association for the Advancement of Science.

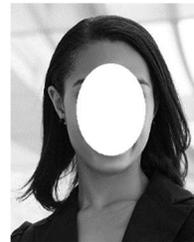

**Daniel V. Klein** has been at Google since 2010, and joined Google DeepMind in 2024 as a Senior Software Engineer. Daniel was VP Technology Development of Lonewolf Systems (1992-2012), Senior Member of Technical Staff at the Software Engineering Institute (1986-1992), Education Director of the USENIX Association (1990-2013), and Manager of Software Systems at the Computer Engineering Center of Mellon Institute (1977-1984). He earned a Masters of Applied Mathematics in 1983 from Carnegie Mellon University, Pittsburgh PA, and BS in Mathematics from Carnegie Mellon University in 1977.

He holds 32 US and 7 foreign patents.

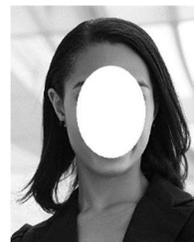

**Theodore L. Ross** earned his Master of Software Engineering degree from Carnegie Mellon University, Pittsburgh PA (Software Engineering Institute) in 1993. This degree was sponsored by Digital Equipment Corporation under the Graduate Engineering Education Program (GEEP). He earned a Bachelor of Science degree in electrical and computer engineering from Carnegie Mellon University in 1986.

He has worked at Red Hat Inc. in Westford and Lowell MA since 2007 as a Senior Principal Software Engineer. He previously worked at Top Layer Networks, Digital Equipment Corporation and did an undergraduate internship at Sandia National Laboratories in Livermore CA.

Mr. Ross holds 12 US Patents.